\documentclass[11pt,onecolumn,amssymb,nofootinbib]{revtex4}
\usepackage{amsmath, amsthm, amscd, amssymb}
\usepackage{graphicx, braket}
\usepackage{bm}
\usepackage{bbm}
\usepackage{epstopdf}

\begin{document}

\title{\bf Is ``Dark Energy'' a Quantum Vacuum Energy?} \bigskip

\author{Stephen L. Adler}
\email{adler@ias.edu} \affiliation{Institute for Advanced Study,
1 Einstein Drive, Princeton, NJ 08540, USA.}

\begin{abstract}
We review the origins, motivations, and implications for cosmology and black holes, of our proposal that ``dark energy'' is not a quantum vacuum energy, but rather arises from a Weyl scaling invariant nonderivative component of the gravitational action.
\end{abstract}

\maketitle

\section{Introduction}
In a series of papers, Refs. \cite{adler1}--\cite{adler7}, we have proposed an alternative theory for the ``dark energy'' that drives the accelerated expansion of the universe.  In this theory, dark energy is not a vacuum energy, but rather arises from a modification of the usual cosmological constant action  that renders it Weyl scaling invariant.  The aim of this article is to give a concise review of  our work, from the initial proposal and checks for consistency with established observations, through implications of the modified action for late time cosmology and for the physics of black hole horizons, to open problems and observational tests.\footnote{ In the papers Refs. \cite{adler1}--\cite{adler6} we used a $(+,-,-,-)$ metric convention; in the papers Refs. \cite{adler7}, \cite{adler8} we used a $(-,+,+,+)$ metric convention. So $ {}^{(4)}g\equiv - {\rm Det} (g_{\mu\nu})$ is positive in both conventions.      We follow the originally used conventions in discussing the respective sets of papers. In Eqs. \eqref{fine1} and \eqref{fine2}, we adopt the convention of setting the Planck constant $\hbar$ and the velocity of light $c$ to unity, $\hbar=c=1$.  In Eqs. \eqref{AB} and \eqref{est} and Sec. VIIIA we set the Newton constant $G$ to unity, $G=1$. }

\section{The conventional picture of ``dark energy''}

An important development in observational cosmology over the last two decades has been the discovery   that the expansion of the universe is accelerating.\cite{accel}  This is conventionally interpreted as evidence for a cosmological ``vacuum energy'' term in the gravitational action of the form,
\begin{equation}\label{cosm}
S_{\rm cosm}=-\frac{\Lambda}{8 \pi G} \int d^4x ({}^{(4)}g)^{1/2}~~~,
\end{equation}
with the cosmological constant $\Lambda$ given by $\Lambda=3H_0^2 \Omega_{\Lambda} $ in terms of the Hubble constant $H_0$ and the
cosmological fraction $\Omega_{\Lambda}$. This functional form incorporates the usual assumption that gravitational physics, including the dark energy that leads to accelerated expansion of the universe, is four-space general coordinate invariant, with no frame dependence in the fundamental
action.  The action assumed in Eq. \eqref{cosm} goes back to the inception of general relativity, when Einstein introduced a cosmological constant term in the gravitational action, only to soon remove it when the Hubble expansion was discovered (but was believed to not accelerate).\cite{cosm}

Although the existence of a small acceleration of the expansion of the universe is now firmly established,\cite{accel} adoption of the action of Eq. \eqref{cosm} to describe it leads to a conundrum known over the years as the ``cosmological constant problem'', or more accurately, the cosmological constant ``fine tuning problem''.  As written, Eq. \eqref{cosm} interprets $\Lambda$ as a vacuum energy density,\cite{haber}
\begin{equation}\label{fine1}
\rho_{\rm vac}^{\Lambda} = \frac{\Lambda}{8\pi G} \simeq  (2 \times 10^{-3} {\rm eV})^4~~~,
\end{equation}
with $G$ the Newton gravitational constant.  However,  quantum  zero point energies of bosonic and fermion fields at the Planck scale correspond to
a vacuum fluctuation energy of order
\begin{equation}\label{fine2}
\rho_{\rm vac}^{\rm QM} \sim M_{\rm Planck}^4 = G^{-2} \simeq ( 10^{28} {\rm eV})^4~~~,
\end{equation}
so that the ratio of the observed cosmological constant, interpreted as a vacuum energy density, to the Planck scale quantum fluctuation energy
density, is
\begin{equation}\label{fine3}
\frac{\rho_{\rm vac}^{\Lambda}}{\rho_{\rm vac}^{\rm QM} }\simeq   10^{-123}~~~.
\end{equation}
To achieve such a small value, a tremendous fine tuning is needed, in order for particle vacuum energies to cancel down to the magnitude of $\rho^{\Lambda}_{\rm vac}$.  Even if one assumes that there is a (not yet observed) supersymmetry in fundamental physics broken only at the TeV scale, so that bosonic and fermionic vacuum energies cancel to the
TeV level, a fine tuning of order $10^{-60}$ is still needed if the observed cosmological constant arises as a vacuum energy density.

Many ideas to explain this fine tuning have been explored in the literature.  Perhaps the simplest is to assume that the observed dark energy is {\it not} a vacuum energy, but arises from new physics not interpretable as a vacuum energy.  In this case one still has to understand the small
 numerator $\sim(10^{-3} {\rm eV})^4$ in Eq. \eqref{fine1}, which would be yet another example of the ``hierarchy problem'' presented by the smallness of the known elementary particle masses compared to the Planck mass.  This hierarchy may arise from a dynamical symmetry breaking chain,  or from instanton nonperturbative effects.   However, the argument for a fine tuned cancellation of vacuum energies is undermined  if $\Lambda$ does not arise as a vacuum energy remnant.  It may be that the  fundamental physics theory that underlies both general relativity and
quantum field theory has a selection rule that enforces
{\it exact}  cancellation of bosonic and fermionic vacuum energies to give zero for the coefficient in Eq. \eqref{cosm}, with dark energy arising from an action of a different form. This is the avenue that we have pursued in our series of articles, and that we will explore in the review that follows.\footnote{After the initial posting of this article, I received several emails with comments and references relevant to the preceding discussion.  Steven Bass notes that the fine tuning estimate of Eq. \eqref{fine3} uses a noncovariant cutoff; a covariant dimesional regularization cutoff reduces the fine tuning estimate to  $O(10^{-56})$, with the dominant contribution coming from the Higgs.\cite{bass1}--\cite{bass3} Naresh Dadhich notes that he has long advocated that the cosmological constant should not be interpreted as a vacuum energy, thus avoiding the fine tuning problem.\cite{dad1}--\cite{dad3}  Leila Graef calls attention to an article giving an infrared fluctuation mechanism leading to dynamical relaxation of the cosmological constant to the currently observed
small value.\cite{graef1}}

\section{Weyl's ``gauge invariance'':  Weyl scaling invariance}
The term ``gauge invariance'' is most frequently used to describe invariance of a theory of spin one fields, either  Abelian or non-Abelian,  under shifts in the vector potential.  However, the term was originally introduced into physics by Hermann Weyl, who proposed that ``Eich Invarianz'', that is invariance under a change of scale, or ``gauge'', might be a local symmetry of general relativity.\cite{gauge1},\cite{gauge2}  Weyl's gauge invariance is implemented by a transformation of the spacetime metric $g_{\mu\nu}(x)$ according to
\begin{align}\label{weyl}
g_{\mu\nu}(x) &\to \lambda^2(x) g_{\mu\nu}(x)~~~,\cr
g^{\mu\nu}(x)  &\to \lambda^{-2}(x) g^{\mu\nu}(x)~~~,\cr
\end{align}
with $\lambda(x)$ a  spacetime scalar function of the coordinate $x$.  When $\lambda(x) =\lambda$, a constant, the transformation of Eq. \eqref{weyl} is termed a global Weyl scaling  transformation.
Since the Einstein-Hilbert action of general relativity, which makes  a multitude of well-verified predictions, is not invariant under Weyl scaling,  Weyl's suggestion was soon abandoned as a possible invariance of interest in gravitational physics.

However, there have been recent revivals.  Forger and R\"omer studied properties of gauge field currents and the energy-momentum tensor in classical field theories.\cite{forger} In the course of this they examined the behavior of the classical ``improved''spin zero, spin one-half, and spin one fields under Weyl scaling of the metric, and corresponding Weyl scalings of the classical fields, of the general form
\begin{equation}\label{fieldscale}
\phi(x) \to \lambda^{p_\phi}(x) \phi(x)~~~,
\end{equation}
 Here  the exponent $p_\phi$ is characteristic of the field $\phi(x)$, with $p_\phi$ taking the value for each  type of field that makes its mass zero action  Weyl scaling invariant.  Drawing on this work, Adler in Ref. \cite{adler1} studied the incorporation of a classical metric into his proposed pre-quantum dynamics,\cite{emerge} leading to a conjecture about the structure of the part of the gravitational action involving no metric derivatives, which will be discussed in detail in the next section.  Independently, 't Hooft wrote a prominent Gravitational Essay  advocating Weyl scaling invariance as ``the missing symmetry component of space and time''.\cite{thooft}

\section{Trace dynamics as motivation for local Weyl invariance of the nonderivative gravitation action}

This section is historical, explaining how we were first led to the postulate of a Weyl scaling invariant, but frame dependent dark energy action.  Readers interested primarily in applications of the postulate to cosmology and astrophysics may skip to the next section.

In Ref. \cite{adler1}  we applied our proposed pre-quantum ``trace dynamics'' of Ref. \cite{emerge} to the classical gravitational metric. Trace dynamics is
the dynamics of matrix-valued fields, using an action function formed from the trace of a Lagrangian constructed as a polynomial in these fields.  Essential use is made of an assumed cyclic invariance of the trace, that is, we assume all fields are ``trace class''.  Varying a trace action automatically gives matrix-valued, or ``operator'', equations of motion, without proceeding via canonical quantization of a classical action.  So trace dynamics gives a more general operator dynamics than quantum field theory, with quantum field dynamics (as obtained by
canonical quantization) a special case, corresponding to field quantities that obey the Heisenberg commutation relations.

The proposal made in Ref. \cite{emerge} is that quantum field theory is recovered from trace dynamics by thermodynamic averaging. That is,
quantum field theory is the thermodynamics of an underlying trace dynamics, obtained by averaging over a canonical ensemble constructed as\footnote{Additional terms in the exponent in $\rho$ not shown in Eq.\eqref{canon}  have the same scale invariance properties as ${\bf H}$; see Ref. \cite{adler1} for details.}
\begin{equation}\label{canon}
\rho \propto e^{-\beta {\bf H}} = e^{-\beta \int d^3x (^{(4)}g)^{1/2} Tr T_0^0(t,\vec x)}~~~,
\end{equation}
with ${\bf H}$ the trace Hamiltonian, and $T_0^0$ the mixed index zero-zero component of the energy momentum tensor.  Applying the analysis of Forger and R\"omer, one sees that ${\bf H}$, as well as the trace Lagrangian ${\bf L}$, is globally Weyl scaling invariant for the case of
massless underlying fields.  Let us now define the induced gravitational action as
\begin{equation}\label{ind1}
{\bf S}[g]_{\rm induced} \propto \int d^4x (^{(4)}g)^{1/2} {\bf L}_{\rm AV}~~~,
\end{equation}
where the subscript AV denotes averaging over the canonical ensemble of Eq. \eqref{canon}.  Then since all quantities used to form Eq. \eqref{ind1} are globally  Weyl scaling invariant, the induced gravitational action inherits this invariance property as well.  Also, since all quantities appearing in Eq. \eqref{ind1} are invariant under three-space general coordinate transformations, the induced gravitational action must have this invariance also.   Moreover, if we
restrict ourselves to the part of the induced action that is independent of derivatives of the metric, global Weyl invariance
implies local Weyl invariance, and corresponds, for diagonal metrics with $g_{0i}=g^{0i}=0$, to the unique functional form
\begin{equation}\label{ind2}
 {\bf S}[g]_{\rm induced} =K \int d^4x (^{(4)}g)^{1/2}  (g_{00})^{-2}~~~,
\end{equation}
with K a constant.  Many more details of the argument just sketched are given in Ref. \cite{adler1}.

In addition to Weyl scaling invariance, and three-space general coordinate invariance, the induced action of Eq. \eqref{ind2} inherits one other key property from the construction of Eq. \eqref{ind1}.  Since the trace Hamiltonian is the time component of a four-vector, the canonical ensemble of Eq. \eqref{canon} picks out a preferred frame, and so therefore does the induced gravitational action.  Thus, the action of Eq. \eqref{ind2} has the unusual property that it is   {\it not} four-space general coordinate invariant. We shall assume, on grounds of simplicity, that its preferred frame is the rest frame of the Cosmological Microwave Background (CMB) radiation. Specifically, the action of Eq. \eqref{ind2} has rotational symmetry,  as does the CMB in its rest frame when small fluctuations are averaged, so we are assuming that there is only one inertial frame with this property, which is both the rest frame of the CMB and the frame to which the
metric component $g_{00}$ in Eq. \eqref{ind2} is referred.

\section{Key Postulate}

Motivated by these arguments, we now introduce our key postulate, that the part of the gravitational action that is independent
of metric derivatives is global Weyl scaling invariant, three-space general coordinate invariant, but frame dependent.
The Einstein-Hilbert action, on the other hand, which involves metric derivatives, is assumed to have its usual non-Weyl-scaling invariant form.    The Weyl scaling invariance condition for the non-derivative part of the gravitational action rules out a
dark energy action of the traditional form of Eq. \eqref{cosm}, that is, it requires that the underlying dynamics leads to a sum rule giving exact
cancellation of bosonic and fermionic vacuum energies.  Instead, identifying the constant $K$ in Eq. \eqref{ind2} with the usual numerical factors  times the
observed cosmological constant, we propose that the dark energy action has the form
\begin{equation}\label{dark}
S_{\rm eff}=-\frac{\Lambda}{8 \pi G} \int d^4x ({}^{(4)}g)^{1/2}(g_{00})^{-2}~~~.
\end{equation}
To set up a phenomenology  to distinguish between $S_{\rm cosm}$ and $S_{\rm eff}$,\cite{adler4} we form an action $S_{\Lambda}$ that is a linear combination of
the two,
\begin{equation}\label{phen}
S_{\Lambda} = (1-f) S_{\rm cosm}+f S_{\rm eff}= -\frac{\Lambda}{8 \pi G} \int d^4x ({}^{(4)}g)^{1/2}[1-f+f(g_{00})^{-2}]~~~,
\end{equation}
so that for $f=0$ one has only $S_{\rm cosm}$, for $f=1$ one only has $S_{\rm eff}$, and for $g_{00}\equiv 1$ the parameter $f$ drops out of Eq. \eqref{phen}. Thus, $S_{\Lambda}$ depends on $f$ only through the deviation of $g_{00}$ from unity arising from metric perturbations.  Setting up the calculation this way permits a convenient check that metric perturbation formulas reduce to the standard ones when $f=0$. In Sec. VII will only quote final results specialized to $f=1$.\footnote{In Sec. 4 of Ref. \cite{adler6}, we made interpretive statements about  $0<f<1$, which we now think should be ignored.  For $f=0$,  metric perturbations that depend only on $t$ can be absorbed into the definitions  of the proper time  $\tau$ and the expansion parameter $a(\tau)$, so nothing new happens. For an action with $g_{00}^{-2}$ replaced by $g_{00}^{-p}$, working to first order in metric perturbations $f$ enters through the product $fp$.}  For $f=1$ one cannot eliminate frame-dependent effects by redefining the time variable, since as shown in Sec. IIA of Ref. \cite{adler5}, and is explicitly seen in Eqs.  \eqref{gencase}--\eqref{gener1}   below, one is prevented from doing so by the fact that Eq. \eqref{phen} is only three-space, and not four-space general coordinate invariant.  Hence observable new physics effects can arise for $f=1$.

\section{Objections and Worries}

Years ago, when attending a Stanford Linear Accelerator Center program committee meeting, we heard one of the participants make the memorable statement: ``Every good experiment needs a first class worrier''. In other words, if one doesn't worry about everything that might have gone wrong, you are apt to fool yourself and get spurious results.  The same holds true for novel theoretical proposals.  In this section we go over various objections to, and worries about, the proposed dark energy action of Eq. \eqref{dark}.

\begin{enumerate}

\item {\bf Q} Doesn't Eq. \eqref{dark}  conflict with the relativity principle?   {\bf A}  According to special relativity, the laws of physics take the same form in all ``inertial'' frames moving uniformly with respect to one another.  This used to be interpreted as a statement that one cannot specify a distinguished inertial frame.  However, observation of the dipole component of the CMB does allow one to determine one's relative velocity with respect to the CMB rest frame, so there now is a way to specify the absolute velocity of an inertial frame: just build a radiometer and make an all sky map of the CMB.   Given that a preferred frame exists, this opens the possibility that low level physical effects may be linked to that frame.   As already noted, since the CMB is isotropic  in its rest frame  (apart from very small fluctuations), it is natural to identify the preferred frame of Eq. \eqref{dark}, which has spherical symmetry, with the CMB rest frame.

\item  {\bf Q} Isn't Eq. \eqref{dark} ruled out by the accurate verification of the standard model of cosmology?  {\bf A} The standard homogeneous, isotropic Friedmann-Lema\^itre-Robertson-Walker (FLRW) line element has the form $(ds)^2=(dt)^2-a^2(t)\big((dr)^2 + r^2 d\Omega^2\big)$, with $a(t)$ the expansion factor.  Thus since $g_{00}=1$ for this line element, the dark energy actions of Eqs. \eqref{cosm} and \eqref{dark} are indistinguishable in FLRW cosmology.  Changes due to $f\neq 0$ become apparent only at first order in perturbations from uniformity, and will be discussed in
    Sec. VII.

\item  {\bf Q}  The standard derivation of a conserved energy-momentum tensor,\cite{wein} that is used to form Einstein equations which are consistent with the Bianchi identities, relies on four-space general coordinate invariance.  So how can one get consistent Einstein equations with the dark energy action of Eq. \eqref{dark}?  {\bf A}  One adopts  a 3+1 point of view,\cite{adler1} as is done in formulating the initial value problem in general relativity.\cite{adm}   Varying the total action, obtained by adding Eq. \eqref{dark} to the Einstein-Hilbert action, with respect to the spatial components $g_{ij}$ of the metric, gives the spatial components of the Einstein equations, including spatial components of the energy-momentum tensor $T^{ij}$.  The remaining components $T^{00}$ and $T^{0i}=T^{i0}$ are then fixed by imposing covariant conservation on $T^{\mu\nu}$, or equivalently, by using Bianchi identities for the Einstein tensor $G^{\mu\nu}$.  For spherical and axial black holes we find that constructing a ``covariant completion'' of $T^{ij}$ in this manner requires only solving algebraic equations,\cite{adler2},\cite{adler7} while for cosmological metrics only an ordinary differential equation with respect to the time variable needs to be solved.\cite{adler4},\cite{adler5}

\item {\bf Q}  Doesn't Eq. \eqref{dark} permit scalar gravitational waves, which are ruled out observationally?  The usual proof that there are no scalar gravitational waves makes use of four-space general coordinate invariance. {\bf A}  It turns out that just the three-space general coordinate invariance of Eq. \eqref{dark} is enough to eliminate scalar propagating waves, but to see this it is necessary to calculate  the $f\neq 0$ corrections to the theory of first order perturbations around the FLRW metric.  This was done in Ref. \cite{adler4}, where we calculate the changes in the scalar equations obeyed by the perturbation amplitudes $A$, $B$, $E$, and $F$, in the notation of
    Weinberg,\cite{wein1} and where we show that the amplitude $B$ can be eliminated by using  three-space general coordinate invariance.\footnote{The gauge choice $B=0$ could be termed ``semi-Newtonian'' gauge; choosing either full Newtonian gauge ($B=F=0$) or
    synchronous gauge ($E=F=0$) would require making a time-dependent gauge transformation, which is not an invariance of Eq. \eqref{dark}.} Dropping matter source terms, we set up the homogeneous equations for scalar waves, assuming that the expansion factor $a$ and the ratios $\dot{a}/a\equiv H$ and $\ddot{a}/a\equiv H^2Q$ are slowly varying. Making a Fourier Ansatz for a possible scalar wave with spatial dependence\footnote{By choosing the spatial dependence of the scalar wave to include a factor of $a$ in front of the wave number $\vec k$, the slowly varying quantity $a$ factors out of the equations for the Fourier coefficients.}   $e^{i a \vec k \cdot \vec x}$ and time dependence $e^{-i\omega t}$, we solve for the $F$ and $A$ amplitudes in terms of the amplitude $E$ to get the two equations
    \begin{align}\label{scalar}
    0=&E[(\vec k)^{\,2}\alpha(\omega) +\beta(\omega)]~~~,\cr
    0=&E[(\vec k)^{\,2}\alpha(\omega) +\delta(\omega)]~~~,\cr
    \end{align}
    with coefficients
    \begin{align}\label{albet}
    \alpha(\omega)=&(i\Lambda fH/\omega^3)/(1+iH/\omega)~~~,\cr
    \beta(\omega)=&(Q-1)H^2+3i\Lambda fH/\omega~~~,\cr
    \delta(\omega)=&-3\beta(\omega)~~~.\cr
    \end{align}
    Since the two equations for $E$ are inconsistent, we must have $E=0$ and hence no propagating scalar waves.   We note in passing that the perturbation equations in $B=0$ gauge can also be usefully written in terms of $F$ and the two ``gauge invariant'' perturbation amplitudes $\Phi$ and $\Psi$, which we have done in Appendix A of Ref. \cite{adler5}.  From this, one sees that even with $f\neq 0$, the Einstein equations imply the usual perturbative relation $\Psi=\Phi$.
\item {\bf Q} Astronomical  black holes are observed with masses ranging from around 4 up to $10^8$ solar masses. Is Eq. \eqref{dark}, with the unusual factor $g_{00}^{-2}$, which would become infinite at a horizon where $g_{00}=0$, consistent with this? {\bf A} In Ref. \cite{adler2} the Schwarzschild-like spherically symmetric solution, arising from the Einstein--Hilbert action with the addition of the action of Eq. \eqref{dark}, was calculated in a perturbation expansion.  In  terms of polar coordinates with the line element
    \begin{align}\label{schw}
    (ds)^2=&B(r)(dt)^2-A(r)(dr)^2 - r^2 d\Omega^2~~~,\cr
    d\Omega^2=& (d\theta)^2+\sin^2(\theta)(d\phi)^2~~~,\cr
    \end{align}
    we find that near the nominal horizon at $r=2M$ the leading corrections to $A$ and $B$ are
    \begin{align}\label{AB}
    A=&(1-2M/r)^{-1}\left(1+\Lambda \frac{48 M^4}{(r-2M)^2} +O(\Lambda^2)\right)~~~,\cr
    B=&(1-2M/r)\left(1-\Lambda \frac{16 M^4}{(r-2M)^2} +O(\Lambda^2)\right)~~~. \cr
    \end{align}
    The correction terms become significant only for when the distance $|r-2M|$ from the nominal horizon is of order
    \begin{equation}\label{est}
    |r-2M|\sim \Lambda^{1/2} M^2 =\Lambda^{1/2} M_\odot^2 (M/M_\odot)^2 \simeq 2 \times10^{-18} {\rm cm} (M/M_\odot)^2 ~~~,
    \end{equation}
    with $M_\odot$ the solar mass.
    So except very near the nominal horizon, the metric is very close to the standard Schwarzschild black hole metric.  In Ref. \cite{adler2}, the corresponding perturbation expansion was also calculated in isotropic coordinates, with a similar conclusion. We will have more to say about
    implications for black hole structure in Sec. VIII below.
\end{enumerate}

\section{Implications for cosmology}

In the papers Refs. \cite{adler5},\cite{adler6} we used the first order perturbation analysis of Ref. \cite{adler4} to discuss implications of Eq. \eqref{phen} for late time cosmology.  In Ref. \cite{adler5} the analysis was done using coordinate time $t$ , and the results were then reexpressed in terms of proper time $\tau$ by a rescaling of the initially assumed Hubble parameter.  In  Ref. \cite{adler6}, all results were expressed from the outset in terms of proper time and the Hubble parameter  $H_0^{\rm Pl}\simeq 67.27 {\rm km}\rm {s}^{-1} {\rm Mpc}^{-1}$ measured by Planck,\cite{planck} and we follow this presentation here, giving results specialized to $f=1$, corresponding to Eq. \eqref{dark} being the sole dark energy action.

In terms of the proper time $\tau$, the line element for our model takes the form\footnote{We follow the convention of using $[\tau]$ to denote functional dependence on the proper time $\tau$, and $(t)$ to denote functional dependence on the coordinate time $t$, where $t$ is a function $t[\tau]$ of $\tau$, and conversely, $\tau$ is a function $\tau(t)$ of $t$.  Thus, for any time-dependent quantity $F$,
$F(t)=F(t[\tau]) =F[\tau]=F[\tau(t)]=F(t)$, with $F[~]$ and $F(~)$ in general being different functions of the argument, unless $\tau \equiv t$.}
\begin{equation}\label{line}
(ds)^2=(d\tau)^2-\psi^2[\tau] (d{\vec x})^{\,2}~~~,
\end{equation}
with $\psi[\tau]$ given in the matter-dominated era by the following formulas,
\begin{align}\label{psieq}
\hat\Phi(x)\equiv &\Phi(x)/\Phi(0)~~~,\cr
 \Delta(x)\equiv & \Phi(x)-\Phi(0)= \Phi(0)[\hat \Phi(x)-1] ~~~,\cr
\psi[\tau]=&a[\tau]\left[1-\Delta(x)\right]~~~,\cr
a[\tau]=&\left(\frac{\Omega_m}{\Omega_{\Lambda}}\right)^{1/3} \Big(\sinh(\hat x[\tau])\Big)^{2/3}~~~,\cr
\hat x[\tau]=&x-\int_0^x du \Delta(u)~~~,\cr
x=&\frac{3}{2} \surd{\Omega_\Lambda}H_0^{\rm Pl}\tau~~~.\cr
\end{align}

Here $\Omega_m \simeq 0.321$ and $\Omega_\Lambda\simeq 0.679$ are the cosmological matter and dark energy fractions, for which we have used  the
Planck  2018 values.\cite{planck}  The initial value of the metric perturbation $\Phi$ at $\tau=0$, denoted by $\Phi(0)$, is the sole parameter of the model,  and is treated as a small quantity to first order.  Thus our model extends the six parameter $\Lambda{\rm CDM}$ model to a seven parameter model, with $\Phi(0)$ as the seventh parameter.   When $\Phi(0)=0$, and also in the small $x$ limit, the above equations reduce to the standard equations of FLRW cosmology.

The function $\hat\Phi(x)$ gives the time evolution of the metric perturbation, normalized to unity at $\tau=0$, and is given exactly in terms of a hypergeometric function by the following formula,\cite{adler6}
\begin{align}\label{hyper1}
\hat\Phi(x)=&(1-u)^b \,{}_2F_1(a,b,c; u)~~~,\cr
a=&\frac{1}{2}+b~~~,\cr
b=&\frac{1}{3}[2+\surd{7}]~~~,\cr
c=&\frac{11}{6}~~~,\cr
u=&{\rm tanh}^2(x)~~~.\cr
\end{align}
When Eq. \eqref{hyper1} is expanded in powers of $x$, the first three terms, which suffice at present for comparisons with experiment, are
\begin{equation}\label{expan}
\hat\Phi(x)= 1+\hat C x^2 +\hat D x^4 + O(x^6)~~~,
\end{equation}
with
\begin{align}\label{coeffs}
\hat C =& \frac{2}{11}=0.1818~~~,\cr
\hat D =& -\frac{2}{561}=-0.00357~~~.\cr
\end{align}

Using these formulas, various quantities of astrophysical interest can be calculated, as follows.
\begin{enumerate}
\item{\bf Effective redshift} In terms of $a[\tau]$, the redshift $z_{\rm eff}$ in our model, for light emitted at proper time $\tau$ and observed at the present proper time $\tau_0$, is given by
\begin{equation}\label{red}
1+z_{\rm eff}=\frac{\psi[\tau_0]}{\psi[\tau]}=\frac{1}{\psi[\tau]} = \frac{1}{a[\tau][1-\Delta(x)]}~~~.
\end{equation}
\item{\bf Hubble parameter}
Corresponding to the line element of Eq. \eqref{line}, the Hubble parameter $H_{\rm eff}[\tau]$ is given by
\begin{align}\label{hubeff}
H_{\rm eff}[\tau]=&\frac{d\psi[\tau]/d\tau}{\psi[\tau]}\cr
=&\frac{d\psi[\tau]/dx }{\psi[\tau]} ~dx/d\tau\cr
=&\left[\frac{da[\tau]/d\hat x[\tau]}{a[\tau]}~d\hat x[\tau]/dx-\frac{d\Delta(x)}{dx}\right]\frac{3}{2}\surd{\Omega_\Lambda}H_0^{\rm Pl}  \cr
=&H_0^{\rm Pl}\surd{\Omega_\Lambda} {\rm coth}(\hat x[\tau])\left[1-\Delta(x)-\frac{3}{2} {{\rm tanh}(x)}\frac{d\Delta(x)}{dx}
\right]~~~.\cr\end{align}
Dividing by $H_0^ {\rm Pl}$, squaring, and using $\coth^2(\hat x[\tau])=1+1/\sinh^2(\hat x[\tau])$, and then using Eqs. \eqref{psieq} and \eqref{red} to eliminate $\sinh^2(\hat x[\tau])$ in terms of the cube of  $(1+z_{\rm eff})[1-\Delta(x)]$, we get the result
\begin{align}\label{final}
\left(\frac{H_{\rm eff}[\tau]}{H_0^{\rm Pl}}\right)^2=& \tilde{\Omega}_m (1+z_{\rm eff})^3 +\tilde{\Omega}_{\Lambda} ~~~,\cr
\frac{\tilde{\Omega}_m}{\Omega_m}=& 1-5\Delta(x)-3\,{\rm tanh}(x)\frac{d\Delta(x)}{dx}~~~,\cr
\frac{\tilde{\Omega}_\Lambda}{\Omega_\Lambda}=&1-2\Delta(x)-3 \,{\rm tanh}(x)\frac{d\Delta(x)}{dx}~~~,\cr
\end{align}
with $\tilde{\Omega}_m$ and $\tilde{\Omega}_\Lambda$ redshift-dependent effective matter and dark energy densities.
To calculate $x$ in the above formulas from $z_{\rm eff}$, it suffices to use the zeroth order formulas
\begin{align}\label{xtau}
x = &{\rm arcsinh}(s)=\log(s + (s^2+1)^{1/2})~~~,\cr
s=&\left(\frac{\Omega_\Lambda}{\Omega_m}\right)^{1/2}\frac{1}{(1+z_{\rm eff})^{3/2}}~~~.\cr
\end{align}
An interpretation of Eq. \eqref{final}  as a modified dark energy equation of state,
\begin{align}\label{stateq}
w_{\Lambda}(z)=&-1+\frac{1}{3}(1+z)\frac{d}{dz}\log U(z)~~~,\cr
U(z)=&\frac{1}{\Omega_{\Lambda}}\left[\frac{\big(\tilde\Omega_m (1+z)^3+\tilde\Omega_{\Lambda} \big)|_z}{\big(\tilde\Omega_m (1+z)^3+\tilde\Omega_{\Lambda}\big)|_0}-\Omega_m (1+z)^3\right]~~~,\cr
\end{align}
is given in Appendix B of Ref.  \cite{adler6}.

\item{\bf Present proper time $\tau_0$} The present proper time $\tau_0$ is determined by the condition $\psi[\tau_0]=1$. Equivalently, we have to calculate $x_{\tau_0}=(3/2) \surd{\Omega_\Lambda}H_0^{\rm Pl} \tau_0$, and to do this we proceed as follows.  We begin by writing
$\hat x[\tau_0] \equiv x_0+ \Delta \hat x$, where $x_0={\rm arcsinh}\big((\Omega_\Lambda/\Omega_m)^{1/2}\big) \simeq 1.169$,   so that the first order perturbation $\Delta \hat x$ is fixed by
\begin{align}\label{calc1}
[1+\Delta(x_0)]\psi[\tau_0]=&1+\Delta(x_0)=a[\tau_0]\cr
=&\left(\frac{\Omega_m}{\Omega_{\Lambda}}\right)^{1/3} \left(\sinh(\hat x[\tau_0])\right)^{2/3}\cr
=&\left(\frac{\Omega_m}{\Omega_{\Lambda}}\right)^{1/3} \left(\sinh(x_0+\Delta \hat x)\right)^{2/3}\cr
=&1+(2/3)\coth(x_0)\Delta \hat x~~~,\cr
\end{align}
which using $\coth(x_0)=1/\surd{\Omega_\Lambda}$ gives
\begin{equation}\label{calc2}
\Delta \hat x=(3/2) \surd{\Omega_\Lambda} \Delta(x_0)~~~.
\end{equation}
We then invert the relation between $x$ and $\hat x[\tau]$ to give
\begin{align}
x_{\tau_0}=&\hat x[\tau_0]+\int_0^{x_0} du \Delta(u)\cr
=&x_0+(3/2)\surd{\Omega_\Lambda}\Delta(x_0)+\int_0^{x_0} du \Delta(u)~~~.\cr
\end{align}
Multiplying through by $2/(3\surd{\Omega_\Lambda}H_0^{\rm Pl})$, this is equivalent to
\begin{equation}\label{calc3}
\tau_0=\tau_0^{\rm Pl}+
\frac{1}{H_0^{\rm Pl}}\left[\Delta(x_0)+\frac{2}{3\surd{\Omega_\Lambda}}\int_0^{x_0} du \Delta(u)\right]
\end{equation}
with $\tau_0^{\rm Pl}=2x_0/(3\surd{\Omega_\Lambda}H_0^{\rm Pl})=13.83 {\rm Gyr}$.
\item{\bf Present Hubble parameter} Turning now to the calculation of $H_{\rm eff}({\rm present})/H_0^{\rm Pl}$, from Eq. \eqref{hubeff} we get
\begin{equation}\label{hubeff1}
H_{\rm eff}({\rm present})/H_0^{\rm Pl}=
\surd{\Omega_\Lambda} {\rm coth}(x_0+\Delta \hat x)\left[1-\Delta(x_0)
-(3/2)\surd{\Omega_\Lambda}d\Delta(x_0)/dx_0\right]~~~.
\end{equation}
From the expansion
\begin{equation}\label{expans}
{\rm coth}(x_0+\Delta \hat x)=\frac{1}{\surd{\Omega_\Lambda} }[1+(3/2)(\Omega_{\Lambda}-1)\Delta(x_0)]
= \frac{1}{\surd{\Omega_\Lambda} }[1-(3/2)\Omega_m \Delta(x_0)]~~~,
\end{equation}
we get
\begin{equation}\label{hubeff2}
\frac{H_{\rm eff}({\rm present})}{H_0^{\rm Pl}}=1-\left[\left(\frac{3}{2}\Omega_m
+1\right)\Delta(x_0)+\frac{3}{2}\surd{\Omega_\Lambda}\frac{d\Delta(x_0)}{dx_0}\right]~~~.
\end{equation}
Putting in numerical values $\Omega_m=0.321$, $\Omega_{\Lambda}=0.679$ and  $x_0\simeq  1.169$, together with the $f=1$ values    $\hat\Phi(x_0)-1=0.244$ \, and \, $d\hat\Phi(x_0)/dx_0=0.409$, we get
\begin{equation}\label{hubeff3}
\frac{H_{\rm eff}({\rm present})}{H_0^{\rm Pl}}\simeq 1-0.867 \,\Phi(0)~~~.
\end{equation}
The recently much discussed  ``Hubble tension'',\cite{riess},\cite{hubble}  is a mismatch between the value of the Hubble parameter
$H_{\rm eff}({\rm present})$  measured from supernova counts in the recent epoch, and the value $H_0^{\rm Pl}$ inferred from the Planck analysis of CMB fluctuations, which record the state of the universe at the very early time of matter-radiation decoupling.   The supernova value is found to be about ten percent larger than the Planck value, suggesting that the accelerating expansion of the universe may have speeded up.  From Eq. \eqref{hubeff3}, we see that to fit a tension
 of $\frac{H_{\rm eff}({\rm present})}{H_0^{\rm Pl}}\simeq 1.1$ with the scale invariant action of Eq. \eqref{dark} we would need to choose $\Phi(0) \simeq -0.115$ (corresponding to $1+\Phi(0)\simeq 0.885 \simeq 0.9$ as used in Fig. 1).
For this value of $\Phi(0)$, the correction factors $\tilde\Omega_m/\Omega_m$ and $\tilde \Omega_{\Lambda}/\Omega_{\Lambda}$ in Eq. \eqref{final} are greater than unity by roughly 20 to 30 percent.  Thus it will be important to include these in analyses assessing the viability of our model.
\item{\bf Remarks on possible origins of the  Hubble tension}  If the Hubble tension should turn out to be an artifact arising from mis-calibration of the distance ladder, as suggested in Ref.  \cite{feeney}, or if the Hubble tension is real but arises exclusively from early time physics, as advocated in Ref. \cite{knox}, then our analysis is consistent with  $\Phi(0)=0$.  In this case the dark energy action of Eq. \eqref{dark} would be indistinguishable from the standard action of Eq. \eqref{cosm}, via cosmological measurements, through first order in perturbations.   If the reported Hubble tension survives further testing, and has a late time component, then $\Delta(x_0)=\Phi(0) [\hat \Phi(x_0)-1]$
     must be negative.  Approximating $\Delta(x_0)$ by the quadratic term
$ \Phi(0)\frac{2}{11} x_0^2$ in the expansion of $\hat \Phi(x_0)-1$, a negative value of $\Delta(x_0)$ corresponds to $\Phi(0)$  negative, which could be argued to be plausible since $\Phi$ is an analog of the Newtonian gravitational potential.\cite{wein2a},\cite{wein2b}
\item{\bf Scenario for inflation arising from a frame-dependent dark energy action}  So far we have considered $\Phi$ as a small perturbation around the
FLRW metric.  The generic case when $\Phi$ is not small was also considered in Ref. \cite{adler5}.  The general line element in a homogeneous, isotropic, zero spatial curvature universe, in which physics is invariant in form only under three-space general coordinate transformations, is given by
\begin{equation}\label{gener}
(ds)^2=\alpha^2(t) (dt)^2 -\psi^2(t) (d\vec x)^{\,2}~~~.
\end{equation}
Taking Eq. \eqref{dark} for the dark energy action (i.e., setting $f=1$ in Eq. \eqref{phen}), and assuming that the particulate matter energy-momentum tensor has a relativistic perfect fluid form with pressure $p$ and energy density $\rho$, the $ij$ component of the Einstein equations and covariant conservation of the matter stress-energy tensor give respectively the two equations,\cite{adler5}
\begin{align}\label{gencase}
\frac{2\psi(t)[\alpha(t)\ddot \psi(t)-\dot \alpha(t) \dot \psi(t)] + \alpha(t) \dot \psi^2(t)}{\alpha^3(t)}=& \psi^2(t)\left[\frac{\Lambda}{\alpha^4(t)}-8\pi G p(t)\right]~~~,\cr
\frac{d\big(\rho(t) \psi^3(t)\big)}{dt}=&-3p(t)\dot \psi(t) \psi^2(t)~~~,\cr
\end{align}
with the dot denoting a time derivative $d/dt$.  The corresponding $00$ component of the Einstein equations is shown in Ref. \cite{adler5} to be be a consequence of the two equations in Eq. \eqref{gencase}, and so does not give additional information.  When $\alpha=1+
\Phi$ with $\Phi$ small, and a corresponding perturbative assumption is made for $\psi$, Eqs. \eqref{gencase} were shown in Ref. \cite{adler5} to reproduce the perturbative analysis discussed previously.

But let us now study the case where $\alpha$ is not assumed to be near unity.  As a first step, let us rewrite Eqs. \eqref{gener} in terms of the proper time $\tau$ defined by
\begin{equation}\label{taudef}
d\tau=\alpha(t)dt~~~,
\end{equation}
giving after some algebra
\begin{align}\label{gener1}
\frac{2\psi[\tau] \psi^{\prime\prime}[\tau]+(\psi^{\prime}[\tau])^2}{\psi^2[\tau]}=&\frac{\Lambda}{\alpha^4[\tau]}-8\pi G p[\tau]~~~,\cr
(\rho[\tau] \psi^3[\tau])^\prime=&-3 p[\tau] \psi^\prime[\tau] \psi^2[\tau]~~~,\cr
\end{align}
with the prime denoting a proper time derivative $d/d\tau$. The fact that the transformation to proper time does not eliminate the function
$\alpha[\tau]$ from these equations is a direct reflection of the fact that the dark energy term in the  action from which these equations are derived is not four-space general coordinate invariant.  Consider now what would happen if at some very early time $\tau$ the function $\alpha[\tau]$, instead of being close to unity, is very small and nearly constant, say $\alpha[\tau] \simeq \bar \alpha$.  Neglecting the matter pressure term $8 \pi G p[\tau]$,
the $\psi[\tau]$ equation in Eq. \eqref{gener1} is then approximated by
\begin{equation}\label{gener2}
\frac{2\psi[\tau] \psi^{\prime\prime}[\tau]+(\psi^{\prime}[\tau])^2}{\psi^2[\tau]}=\frac{\Lambda}{\bar\alpha^4}~~~,
\end{equation}
which (choosing the positive square root of $H^2$) is solved by
\begin{equation}\label{soln}
\psi[\tau]=e^{H\tau}~,~~~H^2=\frac{\Lambda}{3\bar \alpha^4}~~~.
\end{equation}
This would correspond to a very early ``inflationary'' period for $\tau<0$, with inflation ending as $\alpha[\tau]$ increases to a value close to and (as suggested by the Hubble tension analysis) slightly below unity at $\tau=0$, which we have taken as the beginning of conventional ``Big Bang'' cosmology.  Our perturbative analysis then shows that $\alpha[\tau]=1+\Phi(0)\hat \Phi(x)$ decreases again for $\tau>0$, and is approximated by a downwards parabola centered on $x=0$,
since $\hat \Phi(x)$ is an increasing function of $x$.  A $\tau<0$  nonperturbative increase in $\alpha$, needed to end the early inflation,  would have to occur  by a mechanism for which we do not  have equations,  but which presumably arises from the pre-quantum theory that gives rise  to the scale invariant action of Eq. \eqref{dark}. A sketch of this suggested behavior is given in Fig. 1.  Note that if  $\Phi(0)$ were $\geq 0$, the downwards parabola of Fig. 1 would be replaced by a straight line $\big(\Phi(0)=0\big)$ or an upwards parabola $\big(\Phi(0)>0\big)$, and a smooth match of the
$\alpha[\tau]$ curve to an inflationary era would not be possible. Thus, in this interpretation, the current Hubble tension, corresponding to $\Phi(0)<0$, is a remnant in the ``Big Bang'' epoch of the inflationary era when $\alpha[\tau]<<1$.
 \begin{figure}[t]
\begin{centering}
\includegraphics[natwidth=\textwidth,natheight=300,scale=1.0]{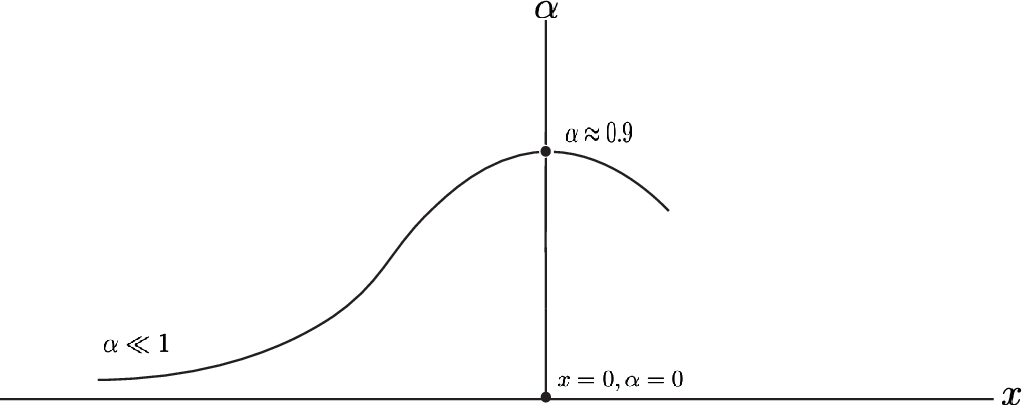}
\caption{Sketch of the  suggested behavior of $\alpha[\tau]$ versus the rescaled proper time $x=\frac{3}{2} \surd{\Omega_\Lambda}H_0^{\rm Pl}\tau$.}
\end{centering}
\end{figure}

In the usual picture of inflation, the role of the  term $\Lambda/\alpha[\tau]^4$ in the above discussion is played by a very large potential energy $V\big(\phi(\tau)\big)$ of a scalar ``inflaton'' field $\phi[\tau]$, with the potential constructed to be very flat near the maximum, with $\phi$ slowly rolling down the potential hill so that $V$ ends up, in the ``Big Bang'' epoch,  being comparable in magnitude to the pressure term $8 \pi G p$. This model for inflation has a huge literature and is well described in standard texts.\cite{wein3},\cite{muk}  Whether an inflaton and its potential are actually
present in the universe, or whether they are  useful proxies for a new type of gravitational dynamics as suggested here, is presently not known.\footnote{
Our interpretation suggests an intriguing calculation. Writing $\Lambda=3 H_0^2 \Omega_\Lambda$ using Planck values, and taking the square root, we get
$H_0^{\rm Pl}=(\Lambda/3)^{1/2}/\surd{\Omega_\Lambda}$, with $\surd{\Omega_\Lambda} \simeq 0.824$.  Extrapolating Eq. \eqref{soln} by defining  a $\tau$ dependent $H[\tau]$ expressed in terms of $\alpha[\tau]$ by $H[\tau]=(\Lambda/3)^{1/2}/\alpha^2[\tau]$, and using the value $\alpha[0]\simeq 0.885$ obtained from fitting the Hubble tension, we get $\alpha^2[0]\simeq 0.783$, which differs by about five percent from $\surd{\Omega_\Lambda}$.  So the extrapolated $H[0]$ is close to the Planck value of the Hubble constant.  This curious observation can be rephrased in a different way: if we impose that the extrapolated $H[0]$ should be equal to $H_0^{\rm Pl}$, we deduce the
relation $0.885 \simeq \alpha[0]\simeq (\Omega_\Lambda)^{1/4}\simeq 0.907$, which is good to a few percent.  Significant, or numerology?}
\footnote{Dabholkar has proposed  imposing Weyl scaling invariance on the total gravitational action, including the
Einstein-Hilbert action, by including an additional ``compensator field'' $\Omega$ which he suggests could play the role of the inflaton.\cite{dab}
This proposal differs from ours, in which the Einstein-Hilbert action retains its original, Weyl scaling breaking form.   }
\end{enumerate}
\section{Implications for black holes}
\subsection{\it Is there an event horizon?}
We turn finally to the implications of the action of Eq. \eqref{dark} for black holes.  We have already noted that a perturbation expansion in powers of $\Lambda$ shows that for the spherically symmetric Schwarzschild-like case studied in Ref. \cite{adler2}, the black hole metric is modified only in a thin shell very close to the nominal horizon.  We turn now to analytic and numerical results obtained in Ref. \cite{adler2}, which show that the $g_{00}^{-2}$ factor in Eq. \eqref{dark} prevents an event horizon from forming.

Heuristically, an event horizon forms when  $g_{00}$ vanishes, so that outgoing light is infinitely redshifted and cannot escape to give an
observable wave at infinity. Similarly, for massive particles the locally measured energy is governed by the equation $E_{\rm local}={\rm constant}/|g_{00}|^{1/2}$,\cite{mtw}  so a particle climbing out from a horizon has its energy infinitely red shifted, and cannot escape to infinity.    Having a $g_{00}^{-2}$ factor in the action, which would become infinite at $g_{00}=0$, changes the equations governing black hole structure so as to prevent the vanishing of $g_{00}$, which leads to absence of an event horizon, and as we shall see later, also absence of an apparent horizon.  When there is no horizon, radiation and matter inside the hole are no longer prevented from climbing out to infinity by an effectively infinite gravitational potential well, leading to the possibility of a flux of radiation and matter emanating from
the hole.

To verify these heuristics, calculations were done in both polar coordinates and isotropic coordinates, with results as follows.
\begin{enumerate}
\item {\bf Polar coordinates}  The static, spherically symmetric line element in polar coordinates is given above in Eq. \eqref{schw}.  When Eq. \eqref{dark} is included in the action, the Einstein equations for $A$ and $B$ can be combined to give a second-order, noninear differential equation for $B$.  To analyze this equation, it is convenient to put it in dimensionless form by scaling out the
    dimensional constant $\Lambda$, by defining\footnote{This rescaled radial variable $x$ (from Ref. \cite{adler2}) is not to be confused with the rescaled proper time variable $x$ of Eq. \eqref{psieq} (from Ref. \cite{adler5}).}
    \begin{align}\label{dimless}
    x=&\Lambda^{1/2}r ~~~,\cr
    B(r)=&B\big(x/\Lambda^{1/2}\big) \equiv b(x)~~~.\cr
    \end{align}
    We then find that $b(x)$ satisfies the differential equation, with ${}^{\prime}$ denoting $d/dx$,
    \begin{equation}\label{master}
    b^{\prime\prime} + \frac{2}{x} b^{\prime} + \frac{2 (x b^{\prime}+b)( x b^{\prime}+3b)}{b(b^2-x^2)}=0~~~.
    \end{equation}
    Analysis of this equation shows that: (i) $b$ cannot vanish, and also cannot become infinite, in the interval $0<x<\infty$;
    (ii) near $x=\infty$, $b$ behaves as a positive constant times $x^{-6}$, corresponding to a singularity at a proper radius of order
    the inverse Hubble constant, where the static assumption is expected to break down;  (iii) moving in from infinity, there is a Schwarzschild-like
    regime $1>>x>>\hat{M}$, where $b\simeq 1-2\hat{M}/x$,  with $\hat M = \Lambda^ {1/2}M$ the rescaled black hole mass; (iv) moving further inwards, the perturbative corrections to $b$ become large, and prevent $b$ from vanishing, so that instead as $b^2$ becomes close to
$x^2$ in size, $b$ develops a branch cut $b(x)=x+C(x-\bar a)^{1/2}$, and becomes complex for $x<\bar a$; (v) below the branch cut $|b|$ becomes infinite as $x^{-1}$ when $x \to 0$, corresponding to the central black hole singularity; (vi)  the cusp-like branch cut is a coordinate singularity, and does not appear in the scalars $R$ (which vanishes identically),  $R_{\mu\nu}R^{\mu\nu}$, and $R_{\mu\nu\lambda\sigma}R^{\mu\nu\lambda\sigma}$.  A sketch of the behavior (i)--(vi)  is given in Fig. 1 of Ref.  \cite{adler2}, followed by plots that illustrate these features in a numerical calculation for $\hat M=10^{-2}$.  {\bf Thus, there is no event horizon where $b(x)=B(r)$ vanishes!}

 \item{\bf Isotropic coordinates} The static, spherically symmetric line element in isotropic coordinates is
\begin{equation}\label{iso}
(ds)^2=\frac{B[r]^2}{A[r]^2} (dt)^2-\frac{A[r]^4}{r^4}[(dr)^2+r^2 d\Omega^2]~~~,
\end{equation}
 where the use of square brackets is a reminder that $A,\,B$  are different functions of a different radial variable than in the polar coordinate case.  When Eq. \eqref{dark} is included in the action, the Einstein equations yield a system of coupled second order differential equations for $A$ and $B$,
 \begin{align}\label{coupled}
 A^{\prime\prime}= & \frac{3}{4}\Lambda \frac{A^9}{r^4 B^4}~~~,\cr
 B^{\prime\prime}=&-\frac{9}{4}\Lambda \frac{A^8}{r^4B^3}~~~.\cr
 \end{align}
 In Ref. \cite{adler2} these equations were developed in a perturbation expansion around the standard isotropic Schwarzschild solution $A[r]=r+M/2~,\,B[r]=r-M/2$, with results analogous to those found using polar coordinates. Additionally, rescaling $A,\,B,\,r$ and $M$ by a factor
 $\Lambda^{1/2}$, the equations of Eq. \eqref{coupled}  were integrated numerically by using the Mathematica integrator, showing small $r$ and large $r$ singularities corresponding to those found in polar coordinates. The numerical  solution is smooth in between the two singularities, with $g_{00}$ real and nonvanishing, in agreement with the polar coordinate results that there is no event horizon, and that the cusp found in polar coordinates is a coordinate singularity, not a physical singularity.
\end{enumerate}

\subsection{\it Is there an apparent horizon?}

An event horizon is a property of the global causal structure of spacetime, which is not accessible in approximate numerical calculations.  So in computational general relativity one employs the concept of the {\it apparent horizon}, defined as the outermost marginally trapped surface.\cite{hawking},\cite{adler8}--\cite{krishnan2}  To see if a compact, orientable  2-surface embedded in 4-space is an apparent horizon, one picks two orthogonal directions corresponding to outgoing and ingoing null rays, with respective tangents $\ell^{\nu}$ and $n^{\nu}$, where $\ell^2=n^2=0$. Using the $(-,+,+,+)$ metric convention, let us fix the inner product of the null rays by adopting the convenient normalization convention,\cite{nielsen}
\begin{equation}\label{normconv}
\ell^{\alpha}n_{\alpha}=-2~~~.
\end{equation}
This  normalization convention  is invariant under reciprocal rescalings of $\ell^{\nu}$ and $n^{\nu}$,
\begin{equation}\label{rescaling}
\ell^\nu \to \kappa \ell^\nu~,~~~n^\nu \to \kappa^{-1} n^\nu~~~,
\end{equation}
with $\kappa=\kappa(x)$ a general
nonconstant scalar function of the spacetime coordinate $x$.

The {\it expansion}  $\theta_{\ell}~(\theta_n)$ of the bundle of null rays associated with the tangent vector $\ell~(n)$  is a measure of the fractional change of the cross sectional area of the bundle as one moves along the central ray of the bundle.
In Ref. \cite{adler8} we showed that the
expansions
$\theta_{\ell}$, $\theta_n$ transform under Eq. \eqref{rescaling} by the simple scaling formulas
\begin{equation}\label{thetatrans1}
\theta_\ell \to \kappa \theta_\ell~,~~\theta_n \to \kappa^{-1} \theta_n~,~~\theta_\ell \theta_n \to \theta_\ell \theta_n~~~.
\end{equation}
Thus calculations of the expansions $\theta_\ell$, $\theta_n$  have a covariance group under the transformations of Eq. \eqref{rescaling},   with the product $\theta_\ell \theta_n$ an invariant.

Consider now what happens if we compute the expansions $\theta_{\ell,n}$ for the same physics viewed from different choices of coordinates.
In each coordinate system we have to pick null vectors $\ell$ and $n$, and different ways of doing this that satisfy the norm convention of
Eq. \eqref{normconv} will differ by the rescaling freedom of Eq. \eqref{rescaling}.  Thus, if we pick the most convenient definitions of $\ell$ and $n$ in each coordinate system, we will in general get {\it different} values of the expansions $\theta_\ell$, $\theta_n$ in the various coordinate systems, with only the product $\theta_\ell \theta_n$ the same in all systems.\footnote{For an illustration of this in the transformation of a Schwarzschild black hole metric from spherical to Gullstrand--Painlev\'e coordinates, see Ref. \cite{adler8}.}  However,  this product is a useful diagnostic.  According to the usual classification reviewed in Refs. \cite{faraoni}--\cite{nielsen}, $\theta_\ell \theta_n<0$ corresponds to a normal or untrapped surface; $\theta_\ell \theta_n>0$ corresponds to a trapped or antitrapped surface; and $\theta_\ell\theta_n=0$ corresponds to a marginal surface, such as the case $\theta_\ell=0$, $\theta_n>0$ which defines the future apparent horizon.  Thus an apparent horizon can be located by computing $\theta_\ell \theta_n$ in any coordinate system, even though the individual values of $\theta_\ell$ and $\theta_n$ may vary from one coordinate system to another.

In Ref. \cite{adler7} we have applied this method to the Schwarzschild-like black hole, which was computed in isotropic coordinates in Ref. \cite{adler2} and discussed above in Sec. VIII A.  We find that the product $\theta_\ell \theta_n \leq 0$ everywhere between the small $r$ and large $r$ singularities, and so there are no trapped surfaces where this product is positive.  Thus unlike the Schwarzschild black hole,  which has an apparent horizon coinciding with its event horizon,
 the Schwarzschild-like black hole has no event horizon, no apparent horizon, and no trapped surfaces!\footnote{Hence, Schwarzschild-like black holes arising from the dark energy action of Eq. \eqref{dark} are a concrete realization of the extremely compact objects discussed
 in Refs. \cite{eco1},\cite{eco2}.}

\subsection{\it Are black holes leaky?}

For a Schwarzschild-like black hole with no interior trapped surfaces, we expect there to be a leakage of particles from the interior to outside the nominal horizon.  We expect this flux even without invoking quantum tunneling, and since it would be at the classical level the
size could be orders of magnitude larger than the quantum mechanical Hawking radiation flux. So we have suggested in  Ref. \cite{adler7} that emerging from black holes, as modified by the dark energy action of Eq. \eqref{dark}, there will be a ``black hole wind'' of particles with enough velocity to escape to an astronomical distance from the hole.  An important issue for the future will be extending the calculation of the preceding section to include normal matter that has entered the black hole, but is not permanently trapped, and developing the physical concepts needed to calculate the flux of particles emerging from the hole.

If there is a  black hole wind, it could have interesting astrophysical consequences.  We here note three of them.
\begin{itemize}

\item  Observations of the central  black hole in our galaxy show the presence of young stars in its close vicinity.  Lu et al. open their article by stating that:\cite{lua}--\cite{luc}  ``One of the most perplexing problems associated with the supermassive black hole at the center of our Galaxy is the origin of young stars in its close vicinity''.  Similar clusters of young stars are found in the vicinity of  supermassive black holes in nearby galaxies.\cite{lu1a}--\cite{lu1d}   A black hole wind of particles may furnish a mechanism for
    the origin of such stars, both by providing material for their formation, and by providing an outward pressure cushioning nascent stars  against the gravitational tidal forces of the black hole.\footnote{For other perplexing phenomena observed near the center of our galaxy, that could be indicative of a black hole wind, see Ref. \cite{huang}.} This idea could be tested in a phenomenological way, by postulating parameters for the black hole wind (particle type, velocity, flux) and seeing if they can account for young star formation.  We are currently pursuing this idea with a simple one dimensional model,\cite{adler9} in which a flux of particles moves upwards along the $z$ axis with a velocity $v$ distribution $f(v)$ at an initial value $z=z_0$, in the presence of a gravitational potential $V(z)=gz$.  The aim is to see if conditions for star formation can be met near the larger  value $z=z^*$ where the particle velocity has decreased to near zero.\footnote{For a related proposal in which an infalling molecular cloud fragments, see Ref. \cite{bonnell}.  I wish to thank Kyle Singh for bringing this reference to my attention.}

 \item  Another interesting place to look for consequences of a black hole wind is in galaxy formation.  Nearly all galaxies are believed to have a central black hole.  Could this association indicate that black holes  catalyze galaxy formation?\footnote{For observational evidence suggesting that massive black holes may have seeded galaxy formation, see Refs. \cite{vest1},\cite{vest2}.}  For example, if there is a wind of particles emerging from the black hole, it could stabilize infalling matter against collapse into the hole, leading
     to distant regions of over-density and star formation.  Again, this idea could be tested at a phenomenological level, by adding a parameterized black hole wind to simulations modeling galaxy formation.

\item The presence of highly collimated jets that emerge from active galactic nuclei, which are believed to contain supermassive black holes, is a striking feature that has attracted much attention.  The prevailing models for the formation of these jets, in particular the Blandford-Znajek  mechanism,\cite{bland} assume no material (other than the negligible Hawking radiation flux) emerges from the horizon of the central black hole.  If this assumption is incorrect, and there is a black hole wind at the classical physics level, this wind could play a contributory role in the formation of the observed jets.  Thus it will be interesting to solve the rotating black hole analog of the calculations of Ref.  \cite{adler2}, to get the structure of a rapidly rotating Kerr-like black hole, to see if there is evidence for preferential emission of a black hole wind along the directions of the rotation axis.  In Ref.  \cite{adler7} we have set up the equations for solving axially symmetric Einstein equations in the presence of the dark energy action of Eq. \eqref{dark}, and discussed a residual general coordinate invariance of these equations that may make obtaining a numerical solution tricky.

    \end{itemize}

\section{Open problems}
We conclude with a short list of suggestions for future work.
\begin{itemize}
\item Our model for the Hubble parameter has a single parameter $\Phi(0)$, so two or more observational data points are needed to give a test.  Since the evolution function $\hat \Phi(x)$ drops from $\simeq 1.24$ at $z=0$ to $\simeq 1.12$ at $z\simeq 0.35$,\footnote{Recall that increasing $z$ corresponds to decreasing $x$, and $z=0$ corresponds to $x=x_0\simeq 1.169$; see Table I of Ref. \cite{adler5}.}  the model predicts that an accelerated Hubble expansion is  ``late time'' in origin.   Thus the model can be strongly confronted when there is precision data on the Hubble parameter for redshifts in the range $0\leq z<1$.
\item It will also be important to see what our model predicts for the so called ``$\sigma_8$ tension'' arising from comparing CMB predictions for clustering with low-$z$ cluster counts, as has already been done for alternative dark energy models in Refs. \cite{lambiase},  \cite{keeley}.
\item  If black holes have no event horizon, quantum state evolution should be unitary from cosmological distances down to the apparent singularity in the black hole interior, and hence there should be no ``quantum information paradox''.   This deserves a detailed exploration in light of the computational results of  Ref. \cite{adler2}.
\item If black holes have no apparent horizon and no trapped surfaces, the Penrose and subsequent theorems,\cite{penrose1},\cite{penrose2}    giving conditions for collapse to black holes and singularities will have to be generalized.
\item Again, if there is no apparent horizon, the question of the interior physics of a black hole becomes physically relevant, and will have to be studied.  This is important for gaining an {\it a priori} understanding of the postulated ``black hole wind''.
\item We have focused on a particle flux emanating from the hole, but there may also be an electromagnetic radiation flux.  This raises the
questions, can leaky black holes still appear ``black'', and how are black hole shadows as observed by the Event Horizon Telescope (see Ref. \cite{event}) affected by the ``leak''?
\item  If there is a ``black hole wind'', there may be significant implications for galaxy and star formation, which have yet to be explored beyond the simple one dimensional model of Ref. \cite{adler9} currently under study.
\item  Although in Ref. \cite{adler7} we formulated the equations for an axially symmetric rotating ``Kerr-like'' black hole with dark energy action given by
Eq. \eqref{dark}, solving these equations numerically is still an open problem.  How does ``leakiness'' manifest itself when nonzero rotation is present, which is the generic case?  Will the solution for rapidly rotating black holes have consequences of interest for active galactic nucleus jets?
\item The argument we gave for a possible connection between inflation and a scale invariant dark energy offers no dynamics for the potential
$\alpha^{-4}[\tau]$ that is postulated to drive inflation, unlike the inflaton models that offer a detailed picture.  What is the physics driving nonperturbative evolution of $\alpha[\tau]$?\footnote{ We do not now think that the model for a nonperturbative ``second equation'' given in Sec. IVC of Ref. \cite{adler5} gives a satisfactory answer.  In the perturbative limit, the relation $\Psi=\Phi$ gives the needed second equation.}
\item Can a future unification theory explain why explicit Weyl scaling symmetry breaking appears in the Einstein-Hilbert action, which is quadratic (and higher) order in metric derivatives, but not in the non-derivative dark energy  action postulated in Eq. \eqref{dark}?
\item  We have obtained the conserving completion of the space-space components of the dark energy stress energy tensor $T^{ij}$ in special background geometry cases where only an algebraic equation, or an ordinary differential equation, need to be solved.  Can one prove that a conserving completion can always be constructed for a general background geometry?
\item What does a scale invariant dark energy action say about the astrophysics of so called ``methuselah stars'', stars with apparent ages very similar to the age of the universe?  Is there other astrophysics, beyond the ones discussed in this review,  that could be altered if dark energy is given by the action of Eq. \eqref{dark}?
\end{itemize}

\section{Acknowledgements}
I wish to thank Kyle Singh for useful conversations, and to thank Nai Phuong Ong, Fethi Ramazano\v glu, and Kyle Singh for reading the initial draft of this paper and making many useful suggestions that were incorporated into the final version.  Thanks also to Amanda Cenker and Audrey Smerkanich for preparing the figure.


\begin{thebibliography}{99}
\bibitem{adler1} S. L. Adler, {\it Class. Quantum Grav.} {\bf 30}, 195015 (2013), arXiv:1306.0482.
\bibitem{adler2} S. L. Adler and F. M. Ramazano\v glu, {\it Int. J. Mod. Phys. D} {\bf 24}, 1550011 (2015), arXiv:1308.1448.
\bibitem{adler3} S. L. Adler, {\it Int. J. Mod. Phys. D} {\bf 25} 1643001 (2016), arXiv:1605.05217.
\bibitem{adler4} S. L. Adler, {\it Int. J. Mod. Phys. D} {\bf 26},1750159 (2017), arXiv:1704.00388.
\bibitem{adler5} S. L. Adler, {\it Phys. Rev. D} {\bf 100}, 123503 (2019), arXiv:1905.08228.
\bibitem{adler6} S. L. Adler, {\it Int. J. Mod. Phys. D} {\bf 30}, 2150044 (2021), arXiv:2008.07598.
\bibitem{adler7} S. L. Adler, {\it Are Black Holes Leaky?}, arXiv:2107:11816.
\bibitem{accel} https://en.wikipedia.org/wiki/Accelerating\_expansion\_of\_the\_universe.
\bibitem{cosm} https://en.wikipedia.org/wiki/Cosmological\_constant.
\bibitem{haber} H. Haber, {\it The Cosmological Constant Problem}, http://scipp.ucsc.edu/$\sim$haber/ph171/CosmoConstant15.pdf.
\bibitem{bass1} S. D. Bass and J. Krzysiak, {\it Phys. Lett. B} {\bf 803}, 135351 (2020), arXiv:2001.01706.
\bibitem{bass2} S. D. Bass and J. Krzysiak, {\it  Acta Physica Polonica B} {\bf 51}, 1251 (2020), arXiv:2004.05489.  
\bibitem{bass3} S. D. Bass, {\it Emergent gauge symmetries--making symmetry as well as breaking it}, {\it Phil. Trans. Royal Society A} (in press), arXiv:2110.00241 (hep-ph).
\bibitem{dad1} N. Dadhich, {\it On the enigmatic $\Lambda$--a true constant of spacetime},  arXiv:1006.1552.
\bibitem{dad2} N. Dadhich, {\it Int. J. Mod. Phys. D} {\bf 20}, 2739 (2011), arXiv:1105.3396.
\bibitem{dad3} N. Dadhich, {\it Understanding General Relativity afer 100 years: A matter of perspective}, arXiv:1609.02138.
\bibitem{graef1} R. Brandenberger, L. L. Graef, G. Marozzi, and G. P. Vacca, {\it Phys. Rev. D} {\bf 98}, 103523 (2018), arXiv:1807.07494.
\bibitem{gauge1} http://www.scholarpedia.org/article/Gauge\_invariance\#Brief\_history\_of\_gauge\_invariance.
\bibitem{gauge2} https://en.wikipedia.org/wiki/Gauge\_theory.
\bibitem{forger} M. Forger and H R\"omer, {\it Ann. Phys.} {\bf 309}, 306 (2004), arXiv:hep-th/0307199.
\bibitem{emerge} S. L. Adler, {\it Quantum Theory as an Emergent Phenomenon} (Cambridge University Press, Cambridge, 2004).
\bibitem{thooft} G. 't Hooft,  {\it Int. J. Mod. Phys. D} {\bf 24}, 1543001 (2015), arXiv:1410.6675
\bibitem{wein}  S. Weinberg,  {\it Gravitation and Cosmology}, John Wiley \& Sons, New York (1972), p. 363.
\bibitem{adm} R. Arnowitt, S. Deser, and C. W. Misner, {\it The dynamics of general relativity}, in {\it Gravitation: An Introduction to Current Research}, ed. L. Witten (Wiley, New York, 1962).
\bibitem{wein1}  S. Weinberg, {\it Cosmology} (Cambridge University Press, Cambridge, 2008), Chapter 5.
\bibitem{planck} N. Aghanim {\it et al.}, arXiv:1807.06209.
\bibitem{riess}  A. Riess, S. Casertano, W. Yuan, L. M. Macri, and D. Scolnic, {\it ApJ} {\bf 876}, 85 (2019), arXiv:1903.07603.
\bibitem{hubble} L. Verde, T. Treu, and A. Riess, {\it Nature Astronomy} {\bf 3}, 891 (2019), arXiv:1907.10625.
\bibitem{feeney} S. M. Feeney {\it et al.}  {\it Phys. Rev. Lett.} {\bf 122}, 061105 (2019), arXiv:1802.03404.
\bibitem{knox} L. Knox and M. Millea, {\it Phys. Rev. D} {\bf 101}, 043533 (2020), arXiv:1908.03663.
\bibitem{wein2a} S. Weinberg, {\it Cosmology} (Oxford University Press, Oxford,  2008, Eq. (5.3.18).
\bibitem{wein2b} S. Weinberg, {\it Gravitation and Cosmology} (John Wiley \& Sons, N. Y.,1972), Eq. (9.1.57).
\bibitem{wein3} S. Weinberg, {\it Cosmology}  (Oxford University Press, Oxford, 2008), Chapter 4 and Chapter 10.
\bibitem{muk}  V. Mukhanov, {\it Physical Foundations of Cosmology} (Cambridge University Press, Cambridge, 2005), Chapter 5 and Chapter 8.
\bibitem{dab}  A. Dabholkar, {\it Phys. Lett. B} {\bf 760}, 31 (2016), arXiv:1511.05342.
\bibitem{mtw} C. W. Misner, K. S. Thorne, and J. W. Wheeler, {\it Gravitation}  (W. H. Freeman and Company, San Francisco, 1973), Sec. 25.4, p. 659.
\bibitem{hawking}  S.W. Hawking and G. F. R. Ellis, {\it The large scale structure of space-time}, ({\it Cambridge Monographs on Mathematical Physics}  (Cambridge University Press, Cambridge, 1973), p. 320.
\bibitem{adler8}   S. L. Adler, {\it Int. J. Mod. Phys. D} {\bf 142}, 2142023, arXiv:2105.07521.
\bibitem{faraoni} V. Faraoni, {Galaxies} {\bf 1}, 114 (2013), arXiv:gr-qc/1309.4915.
\bibitem{krishnan1} B. Krishnan, {\it Quasi-local black hole horizons}, in {\it Springer Handbook of Spacetime}, eds.  A. Ashtekar and V. Petkov  (Springer Verlag, 2014), pp. 527-555, arXiv:1303.4635.
\bibitem{nielsen} A. B. Nielsen and M. Visser, {\it Class. Quantum Grav.} {\bf 23}, 4637 (2006), arXiv:gr-qc/0510083.
\bibitem{ashtekar} A. Ashtekar, C. Beetle, and J. Lewandowski  {\it Class. Quantum Grav.} {\bf 19}, 1195 (2002), arXiv:gr-qc/0111067.
\bibitem{krishnan2} A. Ashtekar and B. Krishnan,  {\it Living. Rev. Rel.} {\bf 7}:10 (2004), arXiv:gr-qc/0407042.
\bibitem{eco1}  V. Cardoso and P. Pani, {\it Nature Astronomy} {\bf 1}, 586 (2017), arXiv: 1709.01525.
\bibitem{eco2}  V. Cardoso and P. Pani, {\it Living Rev. Relativ.}  {\bf 22}, 4 (2019), arXiv:1904.05363.
\bibitem{lua}  J. R. Lu {\it et al.},  {\it ApJ} {\bf 625}, L51 (2005.
\bibitem{lub}  J.R. Lu {\it et al.}  {\it ApJ} {\bf 690}, 1463 (2009).
\bibitem{luc}  http://blackholes.stardate.org/research/milky-way-star-clusters.php.html.
\bibitem{lu1a}  T. R. Lauer {\it et al.}, {\it AJ}  {\bf 116}, 2263 (1998).
\bibitem{lu1b} R. Bender {\it et al.}, {\it ApJ} {\bf 631}, 280 (2005).
\bibitem{lu1c}  A.C. Seth {\it et al.}, {\it AJ 132}, 2539 (2006).
\bibitem{lu1d}  J. R. Lu {\it et al.}, {\it ApJ} {\bf 764}, 155 (2013).
\bibitem{huang} X. Huang, Q. Yuan, and Y.-Z. Fan, {\it Nature Commun.} {\bf 12}, Article number: 6169 (2021).
\bibitem{adler9}  S. L. Adler and K. Singh,  {\it A One-Dimensional Model for Star Formation Near a ``Leaky'' Black Hole}, work in preparation.
\bibitem{bonnell} I. A. Bonnell and W. K. M. Rice, {\it Science} {\bf 321}, 1060 (2008), arXiv:0810.2723.
\bibitem{vest1} M. Vestergaard, {\it ApJ} {\bf 601}, 676 (2004).
\bibitem{vest2} M. Vestergaard, arXiv:0401430 [astro-ph].
\bibitem{bland} R. D. Blandford and R. L. Znajek, {\it Mon. Not. Roy. Astron. Soc.}  {\bf 179}, 433 (1977).
\bibitem{lambiase} G. Lambiase, S. Mohanty, A. Narang, and P. Parashari, {\it Testing dark energy models in the light of $\sigma_8$ tension}, arXiv:1804.07154.
\bibitem{keeley} R. E. Keeley, S. Joudaki, M. Kaplinghat, and D. Kirkby, {\it Implications of a transition in the dark energy equation of state for the $H_0$ and $\sigma_8$ tensions}, arXiv:1905.10198.
\bibitem{penrose1}  R. Penrose, {\it Phys. Rev. Lett.} {\bf 14}, 57 (1965).
\bibitem{penrose2} J. M. M. Senovilla and D. Garfinkle, {\it Class. and Quantum Grav.} {\bf 32}, 124008 (2015), arXiv:1410.5226.
 \bibitem{event}  https://eventhorizontelescope.org/.
\end{thebibliography}
\end{document}